\documentclass[10pt]{bmc_article}

\usepackage{cite} 
\usepackage{url}  
\usepackage{ifthen}  
\usepackage{multicol}   
\usepackage[utf8]{inputenc} 
\usepackage{algorithm2e}
\usepackage{color}
\usepackage{transparent}
\usepackage{verbatim}
\usepackage{amsbsy,dsfont,amssymb,amsfonts,amsmath}
\usepackage[fleqn,tbtags]{mathtools}
\usepackage[mathscr]{eucal}
\usepackage{graphicx} 
\usepackage{epstopdf}
\usepackage{wrapfig}
\usepackage{psfrag}
\usepackage{sidecap}
\usepackage{subfig}
\usepackage{newfloat} 

\newcommand{\clgroup}{\mathsf{CLGrouping}}

\DeclareFloatingEnvironment[name={Supplementary Figure}]{supfigure}
\newcounter{SupTableCounter}
\newcommand{\suptable}[1]{\refstepcounter{SupTableCounter}Supplementary Table \theSupTableCounter.\label{#1}}

\def\beq{\begin{equation}}
\def\eeq{\end{equation}\noindent}
\def\bfd{{\mathbf d}}
\def\hd{\widehat{d}}
\DeclareMathOperator{\mst}{MST}

\newboolean{publ}

\newenvironment{bmcformat}{\baselineskip20pt\sloppy\setboolean{publ}{false}}{\baselineskip20pt\sloppy}

\begin{document}
\begin{bmcformat}


\title{Unsupervised learning of transcriptional regulatory networks via latent tree graphical models}


\author{Anthony Gitter$^{1,2,\dagger,}$\firstauthor \email{Anthony Gitter - gitter@biostat.wisc.edu},
    Furong Huang$^{3,}$\firstauthor \email{Furong Huang - furongh@uci.edu},
    Ragupathyraj Valluvan$^{3}$ \email{Ragupathyraj Valluvan - rvalluva@uci.edu},
    Ernest Fraenkel$^{2,}$\correspondingauthor \email{Ernest Fraenkel - fraenkel-admin@mit.edu},
    Animashree Anandkumar$^{3,}$\correspondingauthor \email{Animashree Anandkumar - a.anandkumar@uci.edu}}


\address{\iid(1)Microsoft Research, Cambridge, MA, USA\\
    \iid(2)Department of Biological Engineering, Massachusetts Institute of Technology, Cambridge, MA, USA\\
    \iid(3)Department of Electrical Engineering and Computer Science, University of California Irvine, Irvine, CA, USA\\
		\iid(\dagger)Current address: Department of Biostatistics and Medical Informatics,
		University of Wisconsin-Madison, Madison, WI, USA and
		Morgridge Institute for Research, Madison, WI, USA}

\maketitle


\begin{abstract}
Gene expression is a readily-observed quantification of transcriptional activity and cellular state that enables
the recovery of the relationships between regulators and their target genes.
Reconstructing transcriptional regulatory networks from gene expression data is a problem that has
attracted much attention, but previous work often makes the simplifying (but unrealistic)
assumption that regulator activity is represented by mRNA levels.
We use a latent tree graphical model to analyze
gene expression without relying on transcription factor expression as a proxy for
regulator activity.  The latent tree model is a type of Markov random field that includes
both observed gene variables and latent (hidden) variables, which factorize on a Markov tree.  Through efficient unsupervised learning approaches,
we determine which groups of genes are co-regulated by hidden regulators
and the activity levels of those regulators.
Post-processing annotates many of these discovered
latent variables as specific transcription factors or groups of transcription factors.
Other latent variables do not necessarily represent physical regulators but instead reveal hidden structure
in the gene expression such as shared biological function.
We apply the latent tree graphical model to a
yeast stress response dataset.  In addition to novel predictions, such as condition-specific binding of the transcription factor Msn4, our model
recovers many known aspects of the yeast regulatory network.  These include groups of co-regulated
genes, condition-specific regulator activity, and combinatorial regulation among transcription factors.
The latent tree graphical model is a general
approach for analyzing gene expression data that requires no prior knowledge of
which possible regulators exist, regulator activity, or where transcription
factors physically bind.
Consequently, it is promising for studying expression datasets
in species and conditions where these types of information are not available or not reliable.
\end{abstract}


\ifthenelse{\boolean{publ}}{\begin{multicols}{2}}{}


\section*{Introduction}
Genome-wide studies of gene expression continue to be a widely-used technique for investigating biological
processes and systems.  Microarrays have enabled the collection of large gene expression datasets
for well over a decade~\cite{gasch2000genomic}, and steady advances in experimental technologies, most notably RNA sequencing,
have brought corresponding improvements in the quality of such datasets.  Exploring which genes
are activated or repressed in specific cell types or biological conditions can serve as a starting
point for understanding gene function~\cite{eisen1998cluster}.  Beyond individual genes, groups of genes that behave
similarly across diverse conditions can lead to the discovery of common transcriptional regulatory mechanisms,
which provides further insight into the cellular reaction to changing conditions~\cite{segal2003module}.
Transcription factors (TFs) are central regulatory proteins that bind the promoter regions of their target genes and control
the expression of those genes.  Genes regulated by the same TF have expression patterns that are correlated with the regulatory activity of that TF.
TFs are themselves regulated, both transcriptionally and by several additional mechanisms.
Post-transcriptional regulation, such as microRNA binding~\cite{pasquinelli2012micrornas},
decouples mRNA and protein expression levels.  Even after a protein is translated, post-translational modifications
can activate or deactivate a TF, and TFs must be localized to the nucleus in order to regulate their
target genes.  Consequently, a TF can be active in a particular condition without being differentially expressed
and vice versa~\cite{lefebvre2010master}.

Computational strategies for recovering transcriptional regulatory networks have a long, rich history
and have been extensively reviewed~\cite{bansal2007infer,de2010advantages,marbach2012wisdom,chasman_network_2016}.
Most existing methods for inferring regulatory networks or modules require a map of TF-gene
interactions~\cite{bar2003computational,liao2003network,lefebvre2010master,wu2013multi,arrieta-ortiz_experimentally_2015},
depend on gene perturbations~\cite{markowetz_nested_2007,haynes2013mapping,macneil_transcription_2015},
or assume that the mRNA expression levels of the gene that encodes a TF
are representative of the TF's regulatory activity~\cite{segal2003module,margolin2006aracne,
friedman_sparse_2008,huynh_inferring_2010,roy_integrated_2013}.
Although this expression assumption is expedient, it is not accurate.

A preferable approach is to include the
many other regulatory processes that act upon TFs in the computational model as hidden, latent effects.
We propose learning a latent tree probabilistic graphical model as an efficient approach for recovering the
transcriptional regulators from gene co-expression data without relying on TF expression. Probabilistic graphical models represent probability distributions that factor according to a certain graph, termed the Markov graph~\cite{Lauritzen:book}.  Latent tree graphical models involve observed and hidden variables that factorize according to a tree model.  In scenarios where hidden factors affect observed phenomena, latent tree models are capable of recovering intrinsic relationships between the observed phenomena and the latent factors. In addition, they provide a flexible approach for modeling hierarchical dependencies found in gene regulation.  Although general graphical models with cycles are NP-hard to learn~\cite{Karger&Srebro:01SODA}, there exist efficient guaranteed approaches for learning latent tree models. In addition, inferring the values of the latent variables given the gene expression levels is NP-hard in general but becomes computationally tractable on a tree. Moreover, biological datasets suffer from high dimensionality. There are far fewer observed samples than unknown parameters in general graphical models, and learning general models is thus ill-posed in the high-dimensional regime. Latent tree models, on the other hand, can be learned efficiently using far fewer samples than the number of nodes in the model.

In this paper, we employ the approach of Choi et al.~\cite{Choi&etal:10JMLR} for learning a latent tree graphical model. This approach is guaranteed to recover the correct underlying tree when samples are drawn from a latent tree graphical model.  Moreover, the algorithm is unsupervised and does not require knowledge of the tree structure or the number and location of the hidden variables. In addition, the approach has the flexibility to trade off the number of latent variables discovered (i.e., model complexity) with fidelity to the observed data.
This latent tree learning algorithm has been successfully employed for automatic categorization of financial and text data~\cite{Choi&etal:10JMLR}, contextual object recognition~\cite{choi2012context}, human pose estimation~\cite{wang2013beyond}, and tracking dynamic social networks~\cite{ValluvanEtal:12Sunbelt}. Alternative latent variable models such as latent Bayesian networks~\cite{pearl2000causality} or topic ad-mixture models~\cite{AnandkumarEtal:DAGicml12} are in general challenging to learn.  Although some papers have shown promise~\cite{AnandkumarEtal:DAGicml12,anandkumar2014tensor}, their applicability to the biological domain, which is highly noisy and data poor (in terms of the number of experimental samples per gene), is unclear and left for future investigation.

In our biological setting, we use the latent tree graphical model to represent the relationships among gene expression (represented as observed variables), expression regulators (represented as hidden variables), and other unobserved factors, where the Markov tree structure represents the hierarchical relationships.  Our goals include learning which regulators and other hidden factors may control specific genes in different conditions, detecting groups of co-regulated genes (modules), and inferring the functional activity of regulators.  We do not assume that regulators are known \emph{a priori} or that regulator activity is observed in the gene expression data. We instead recover the hidden relationships between gene expression levels, find factors that regulate different subsets of the genes, and interpret these latent variables.

We find that most latent variables correspond to specific transcriptional regulators or groups of regulators that drive gene co-expression.  TFs can be mapped to latent variables as a post-processing step and \emph{de novo} motif discovery can provide information about potential physical regulators when TF-gene interactions are unknown. Inference in the graphical model recovers the activity levels of these TFs in the various biological conditions that were surveyed.  Latent variables that do not match TFs may represent additional unobserved influences on gene expression including environmental factors and higher-order biological processes. These unobserved nodes are akin to industrial sectors discovered when analyzing financial data~\cite{Choi&etal:10JMLR}, article topics found by analysis of text~\cite{Choi&etal:10JMLR}, or subjects in image analysis (Figure~\ref{Fig:LatentTreeCartoon}).  Moreover, the neighbors of a given variable in the Markov tree structure are conditionally independent given the variable. In the biological context, this can be interpreted as the conditional independence of the expression levels of a group of genes that are neighbors of (i.e., controlled by) a common regulator or group of regulators.  These edges between a latent variable and a set of genes are potentially more informative than gene-gene edges for understanding regulatory processes because they guide the search for explanations of why groups of genes are correlated.

\begin{figure}[ht]
  \centering
	\includegraphics[width=3in]{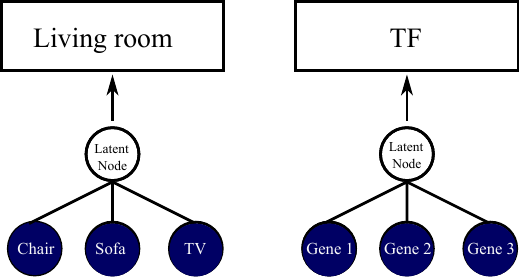}
  \caption{The latent tree algorithm we use to construct transcriptional regulatory networks is analogous to the latent tree approach for discovering objects in images.  In image analysis, co-occurrence of objects in images can be explained by their common dependence on an unobserved `meta object' that is not explicitly labeled in training images. For instance, the labeled objects `Chair', `Sofa', and `TV' are related to the common subject `Living room', which is latent.  Similarly, a latent tree can be used to find groups of genes that are co-expressed across biological conditions due to co-regulation by an unobserved regulator.  The latent tree does not directly provide the regulator-gene interactions, but we demonstrate how external information can be used to reveal the identities of these regulators as specific transcription factors or groups of transcription factors.}
	\label{Fig:LatentTreeCartoon}
\end{figure}

We applied our latent tree graphical model to a compendium of yeast gene expression data covering
various stress and non-stress conditions~\cite{GaschData}.  By studying yeast
we can leverage comprehensive TF binding data~\cite{macisaac2006improved} to annotate
the latent nodes (LNs) and demonstrate how the latent tree recovers well-known yeast stress response
mechanisms.  For other organisms in which TF binding is poorly characterized, we show how
\emph{de novo} motif discovery can be used to identify specific regulators that correspond
to the latent nodes.
In addition, the latent tree predicts pairs of TFs that exhibit combinatorial
regulation in specific stress conditions, and these predictions are supported by independent
data.  To highlight the advantages of not assuming gene expression is a reliable
proxy for TF activity, we compared our results with ARACNE~\cite{margolin2006aracne,margolin2006aracneprot}.  Due to its dependence
on TF expression, ARACNE recovers only a small subset of the important regulators in the
biological conditions we study.
Because the latent tree approach does not require any input data besides gene expression, it
is quite general and can be used to better understand regulatory relationships in many biological settings.

\section*{Results}
We used the latent tree approach to model yeast gene expression levels.  An example of a latent tree is shown in Figure~\ref{Fig:LatentTreeConcept1}.  A variable in the latent tree is conditionally independent of all other variables when conditioned on its immediate neighbors. Thus, the latent tree model provides useful conditional independence relationships between genes' observed expression levels and the discovered latent variables.

\begin{figure}[htp]
\centering
\subfloat[a][\small{Latent tree}]
	{\begin{minipage}{4in}\centering\label{Fig:LatentTreeConcept1}
	\def\svgwidth{4in}
	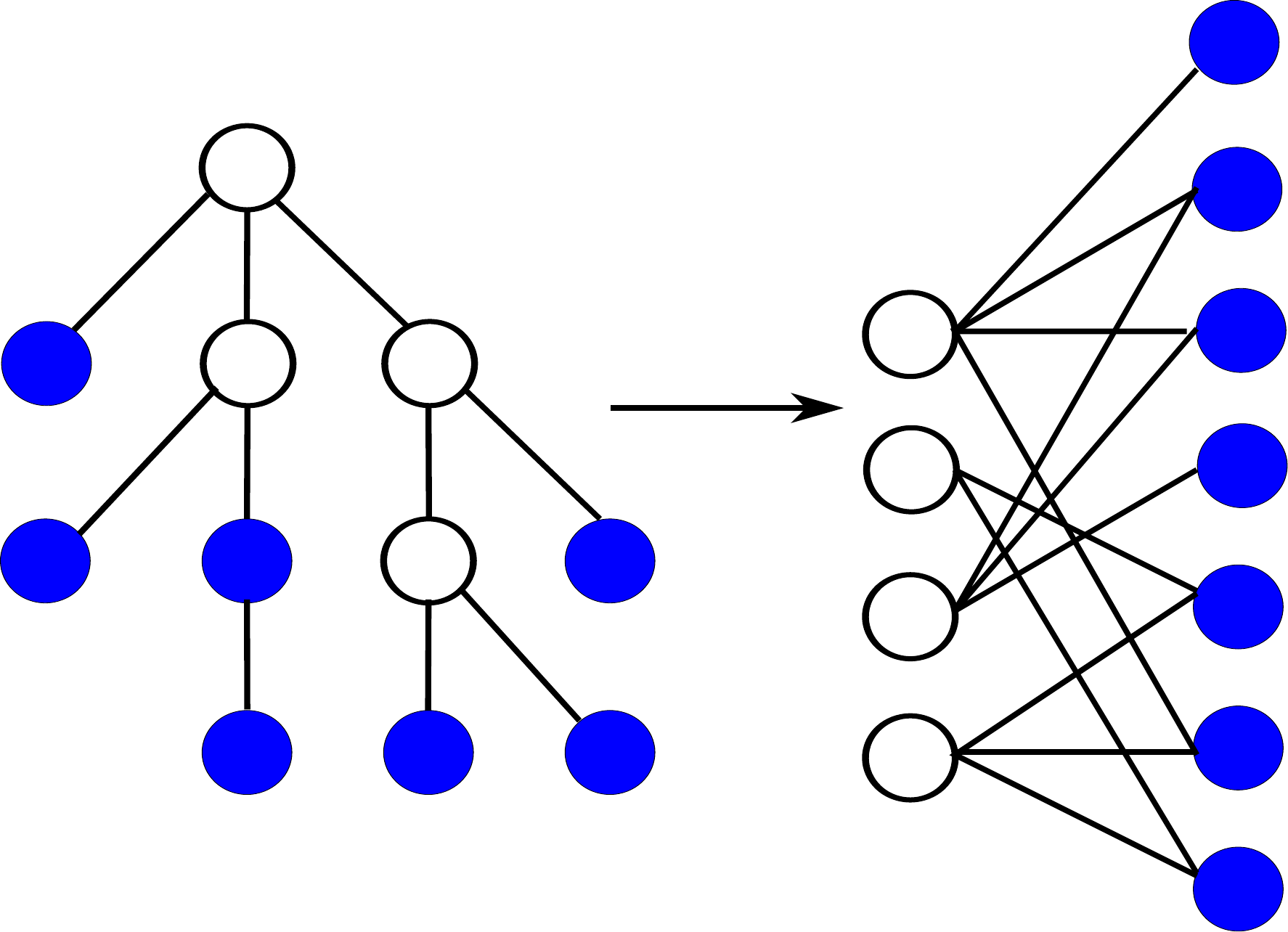
	\end{minipage}}
\hfil
\subfloat[b][\small{TF and GO category annotation}]
	{\begin{minipage}{4.5in}\centering\label{Fig:LatentTreeConcept2}
	\def\svgwidth{4.5in}
	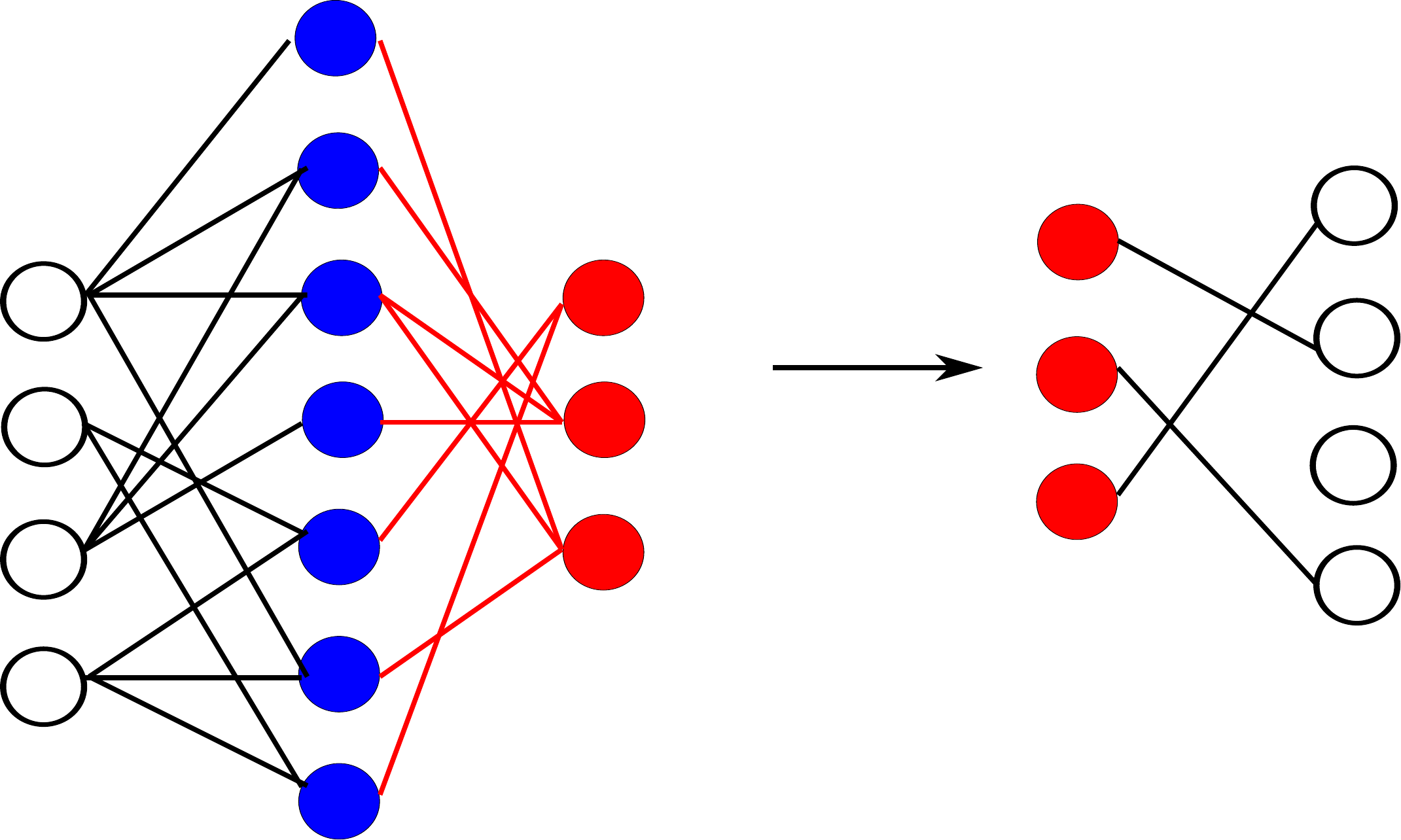
	\end{minipage}}
\caption{An overview of the latent tree approach. (a) We learn a latent tree (left) from the gene expression data, which  introduces latent nodes. $g_i$ denotes an observed gene variable and $y_j$ denotes a latent node.  For each latent node, we define an extended neighborhood of influence that includes all genes that are highly correlated with the latent node activity (right).  The latent node neighborhood may include both direct neighbors in the tree and more distant gene variables, and genes may be included in multiple neighborhoods.  For example, gene $g_5$ is not directly connected to latent node $y_3$ in the latent tree structure (left), but it is influenced by both $y_1$ and $y_3$ (right).  (b) We evaluate the groups of genes influenced by each latent node by annotating the LNs that likely represent specific TFs or Gene Ontology (GO) terms.  We determine which latent nodes may correspond to TFs by assessing the overlap of the LN neighborhoods, the black edges, and the gene targets that are bound by a TF, the red edges (left). Likewise, we search for LN neighborhoods that are enriched for particular GO terms.  Finally, we obtain a bipartite graph representing statistically significant LN-TF pairs and LN-GO term pairs (right). Note that we do not use TF binding interactions or GO annotations when learning the latent tree structure in (a).}
\end{figure}

Our conjecture is that the latent effects that are captured by the latent variables primarily represent the activities of transcription factors.
Latent nodes are introduced naturally by our unsupervised learning method without any knowledge of the TFs.  Figure~\ref{Fig:LatentTreeConcept2} illustrates how we can post-process the latent tree to determine which latent nodes represent known TFs and which are potentially novel regulators.  Specifically, after learning the latent tree, we conduct Fisher's exact test (controlling for multiple hypothesis testing) to find statistically significant relationships between the genes bound by transcription factors and the genes that are correlated with latent nodes.  This general framework allows us to interpret the latent nodes in terms of their relationships with TFs.  We also demonstrate how to annotate latent nodes when TF-gene interactions are unavailable using motif discovery.  Furthermore, we use the TF-latent node mappings to detect potential combinatorial regulation among TFs that are predicted to control the same group of genes.  Latent nodes that do not correspond to specific transcriptional regulators are shown to instead represent biological processes in some cases.

Inference on the latent tree reveals the activity levels (conditional means) of the latent nodes without relying on external TF-gene binding information.  TF activity cannot be directly observed from gene expression data so the inferred values provide a powerful way to detect TFs' context-specific regulatory behaviors.

\subsection*{Modeling Yeast Stress Response}
We applied our latent tree algorithm to a compendium of \emph{Saccharomyces cerevisiae} microarray experiments
composed of many stress conditions (hyperosmotic stress, heat shock, DNA damage, amino acid
starvation, etc.) and normal growth conditions~\cite{GaschData}.  Transcriptional regulation in yeast
stress response has been studied extensively, revealing
the primary TFs that drive transcriptional changes, which allows us to confirm many of our predictions.  Groups of genes that exhibit similar expression profiles
across different stress conditions are in many cases controlled by common TFs, and latent tree analysis of these co-expressed
genes can guide the search for novel stress-specific TF activity.

\subsection*{Latent Tree}
To model unobserved transcriptional processes, we first learn a latent tree network and then determine which genes are likely to be controlled by the hidden regulators in the tree.
We construct a latent tree using $1035$ genes that exhibit substantial expression changes in the yeast stress response data
as the observed variables.  Specifically, we include all genes that have high covariance with at least one other gene.  The latent tree algorithm automatically determines the number of latent nodes using the Bayesian Information Criterion.  However, to improve the biological interpretability of the model and minimize redundant latent nodes, we include a post-processing contraction step that merges latent variables that have a small information distance to an observed gene variable.
Each latent node should reflect a unique biological activity signature, which may be similar
to the signatures of other latent nodes but should not be nearly identical.  In our yeast study, we set the contraction parameter using prior knowledge (Methods) to obtain a latent tree with $90$ latent nodes (Figure~\ref{Fig:LatentTree}a and Supplementary Table~\ref{STab:LatentTreeEdges}).

\begin{figure}[htp]
  \centering
\includegraphics[width=5in]{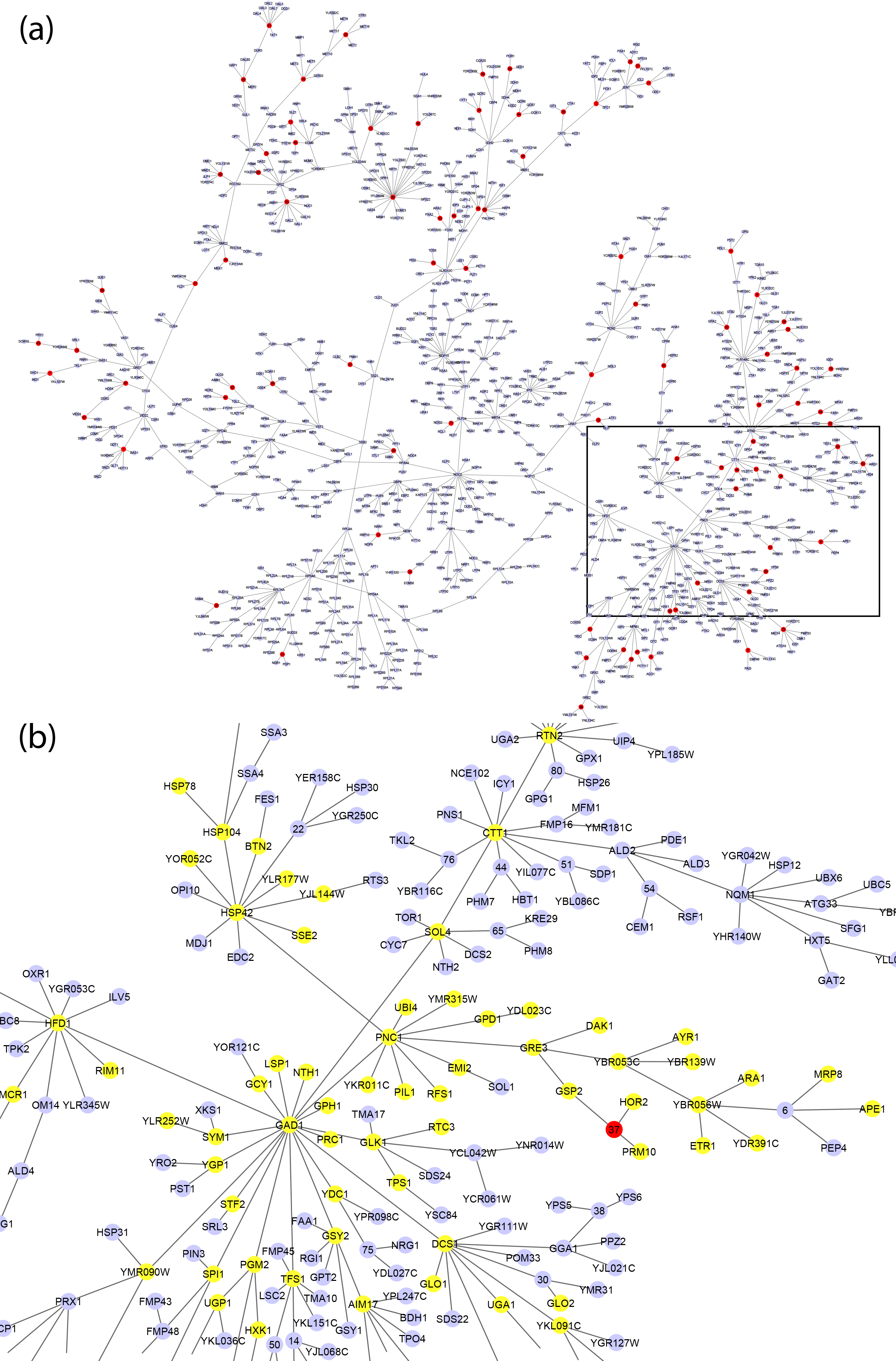}
  \caption{(a) The latent tree structure. Red nodes are the $90$ latent nodes and blue nodes are the observed genes. The black box corresponds to the zoomed portion of the network. (b) Latent node $37$ (red) and its extended neighborhood of influence (yellow).  Most of the latent node's neighborhood is not directly connected to it in the network, but these genes are all highly correlated with latent node $37$'s activity.}
	\label{Fig:LatentTree}
\end{figure}

Supplementary Figure~\ref{SFig:LatentTreeDegree} shows the degree distribution of the LNs.
Fifty LNs have only three direct neighbors, the minimum possible degree of a LN in the latent tree model (Methods).
One hub LN (node $85$) has a degree of $23$ and is connected to many sporulation genes.
The average LN degree is $4.2$, but the influence of the LNs extends beyond their immediate neighbors.
When a group of genes are highly correlated with a LN, only a subset of them are selected by
our algorithm as the direct neighbors of that LN due to the tree constraint on the network structure.
Nevertheless, the LN may control all of these genes and influence
their co-expression (Figure~\ref{Fig:LatentTreeConcept1}).

In order to guide our biological interpretation of the latent tree model, we define an extended neighborhood of influence for each LN (also referred to as the LN neighborhood), which extends beyond
its immediate neighbors in the tree (Supplementary Table~\ref{STab:ExtNeighborhood}).  The LN neighborhood includes
all genes in the latent tree network that are highly correlated with the LN, and each gene may belong to
multiple LN neighborhoods. For instance, latent node $37$, which is associated with osmotic stress response and the TFs Msn2 and Hsf1 (as determined below), is directly connected to only three genes in the latent tree (Figure~\ref{Fig:LatentTree}b) but is likely to regulate all $64$ genes in its extended neighborhood of influence (Figure~\ref{Fig:LN37Neighborhood}).

\begin{figure}[htp]
  \centering
	\includegraphics[width=1.75in]{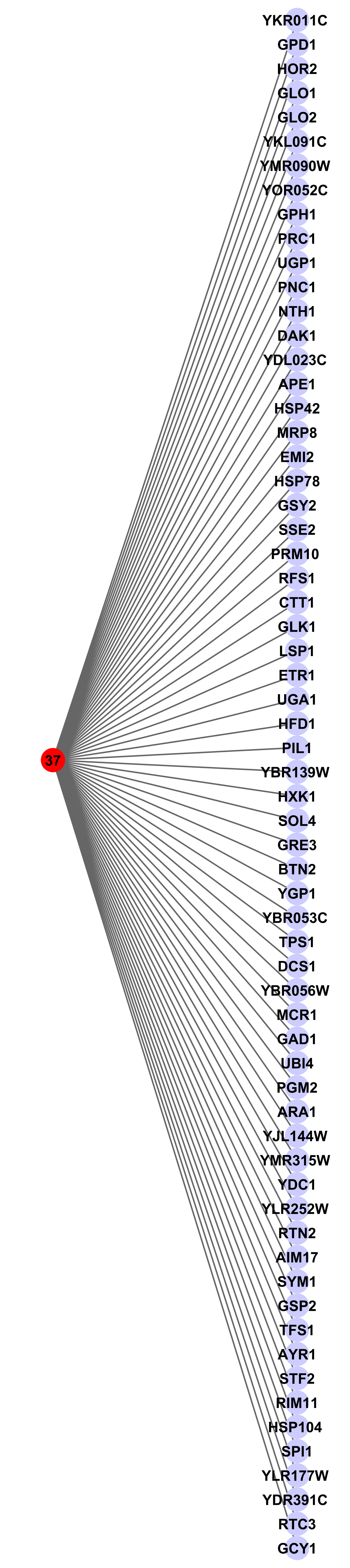}
  \caption{The extended neighborhood of latent node 37, which contains all of the yellow nodes in Figure~\ref{Fig:LatentTree}b.}
	\label{Fig:LN37Neighborhood}
\end{figure}

Recall that the latent tree approach does not use TF-gene binding interactions because our learning approach is unsupervised. 
However, in order to
evaluate our model, we can leverage existing TF-gene binding interactions and annotate which LNs may represent specific TFs or groups of TFs.
We employ a high-confidence yeast TF binding dataset~\cite{macisaac2006improved} to carry out this annotation.
If a LN neighborhood is significantly enriched for genes bound by a specific TF, then it is likely
that the LN signature represents the activity levels of that TF across the biological conditions.  For each LN-TF pair, we compare the LN neighborhood with the set of genes bound by the TF and apply Fisher's exact test to assess the overlap (Figure~\ref{Fig:LatentTreeConcept2}).
Raw $p$-values are calculated for each null hypothesis $\mathcal{H}_{0,ij}$ that the pair $LN_i$ and $TF_j$ are statistically independent.
We use the false discovery rate (FDR) to control for multiple hypothesis testing (Supplementary Methods) and consider a TF and LN to be associated if the FDR is less than $0.05$ (Supplementary Table~\ref{STab:TFSignificance}).  The associated TF-LN pairs can be
represented as a bipartite graph (Figure~\ref{Fig:BipartiteTFLV}). The significant pairs provide a many-to-many mapping between the TFs and LNs and reveal how
specific TFs may induce the correlation structure observed in the gene expression data.

\begin{figure}[t]
  \centering
  \includegraphics[width=\textwidth]{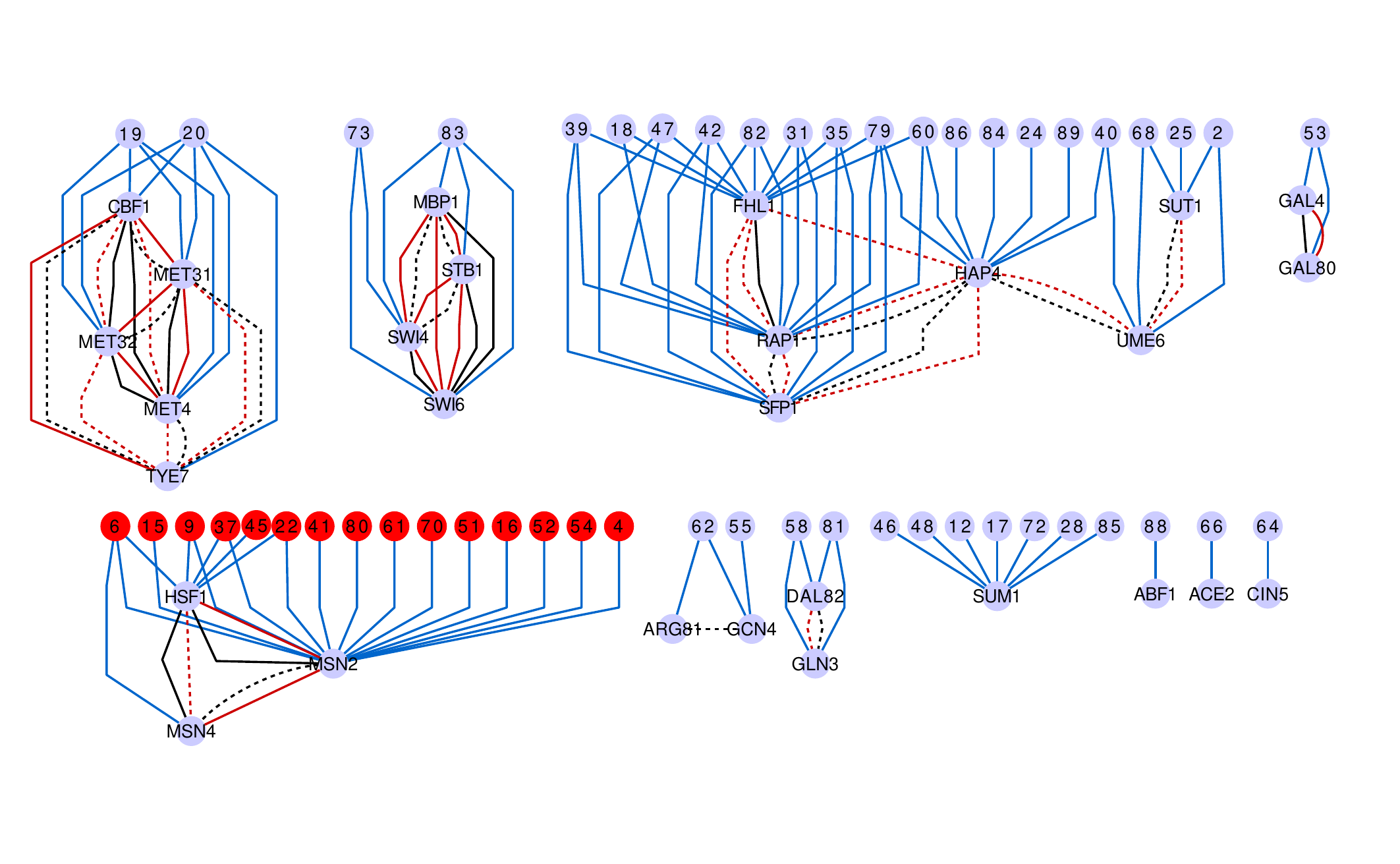}\\
  \caption{Bipartite graph between TFs and LNs (blue edges).  Latent variables are the numbered nodes, where the number is an arbitrarily assigned index.  TF-TF edges are overlaid on the bipartite TF-LN graph and represent known associations between pairs of TFs that are assigned to the same latent node.  Black edges are physical interactions and red edges are genetic interactions.  Solid edges are direct TF-TF interactions and dashed edges signify the TFs interact through an intermediate protein. Red nodes are the latent nodes that are active in osmotic stress response. Other latent nodes that
  are active in this stress condition but not associated with any TFs are omitted.}
	\label{Fig:BipartiteTFLV}
\end{figure}

In our experiments, we find that the majority of the LNs ($51$ out of $90$) are annotated with at least one TF.
On the other hand, only $28$ TFs  match at least one LN.  Many of the samples in the expression dataset are from stress
conditions, and we filter genes that do not vary substantially across the conditions.  These genes perform
other biological functions, and the TFs that primarily bind the removed genes are not expected to appear as predicted regulators
in our latent tree analysis.
TFs that predominantly bind the stress response genes that are included in our latent tree model and are known stress regulators are prevalent among the significant TF-LN pairs, affirming our model's ability to correctly identify regulators of co-expressed genes.  Examples include
Gcn4 in amino acid starvation~\cite{natarajan2001transcriptional},
Hsf1 in heat shock~\cite{hahn2004genome}, Sfp1 in DNA damage~\cite{xu1998sfp1}, and
Msn2 and Msn4 as general stress regulators~\cite{estruch2000stress}.
Other components of the TF-LN bipartite graph capture non-stress biological processes such as the cell cycle
(Mbp1, Stb1, Swi4, and Swi6)~\cite{debruin2008stb1}
and ribosomal protein transcription (Fhl1 and Rap1)~\cite{wade2004transcription} because the genes functioning in these processes
are co-expressed in a sufficient number of the samples.
Latent node $85$, the hub in the latent tree that is connected to many sporulation genes, is associated with
Sum1, a known sporulation gene repressor~\cite{pierce2003sum1}.

Not all latent nodes correspond to transcription factor activity.  In some cases, a latent node can capture
other hidden external effects or represent an entire biological process.  We explore whether the
latent node neighborhoods that do not significantly overlap any TFs could relate to biological processes
by assessing whether they are enriched for
Gene Ontology (GO) terms~\cite{ashburner2000gene} (Table~\ref{tab:GO} and Supplementary Table~\ref{STab:GOSignificance}).
Although these latent nodes may not correspond to specific regulatory mechanisms, they remove direct dependence
among the neighboring genes because the latent tree model posits that
the genes are conditionally independent given the state of the latent process.
Latent node $1$ is one such example.  Its gene neighborhood significantly overlaps with genes annotated with the GO term
`regulation of cell cycle', but these genes do not appear to be under the direct control of the cell cycle
TFs Mbp1, Stb1, Swi4, or Swi6 (Figure~\ref{Fig:BipartiteTFLV}).
This reveals the presence of an additional hidden effect besides the activities of these
four TFs that explains dependencies among a subset of cell cycle genes.
In other cases, the GO terms are associated with latent nodes that do correspond to specific TFs and
complement the known functions of those TFs.
For example, all seven latent nodes associated with Sum1 overlap with the `sporulation' GO term.

\begin{table}[ht]\caption{Statistically significant overlaps between GO biological process terms and latent nodes.}
\centering
\begin{tabular}{c|c}
\hline
GO biological process & Latent nodes\\
\hline
carbohydrate metabolic process (GO:0005975)	 & 4	37	49	52	61	70	78\\
cell wall organization or biogenesis (GO:0071554) &	12	17	28	46	48	72	85\\
cellular amino acid metabolic process (GO:0006520) &	20\\
cellular respiration (GO:0045333) &	24	39	40	60	84	86	89\\
cytoplasmic translation (GO:0002181) &	18	31	35	39	42	47	60	79	82\\
DNA recombination (GO:0006310) &	19	90\\
generation of precursor metabolites and energy (GO:0006091) &	24	40	60	68	84	86	89\\
mitochondrion organization (GO:0007005) &	89\\
nuclear transport (GO:0051169) &	47	82\\
oligosaccharide metabolic process (GO:0009311) &	4	52	61	70\\
organelle assembly (GO:0070925) &		31	34	35	47	82\\
protein folding (GO:0006457) &		45\\
regulation of cell cycle (GO:0051726)&			1\\
ribosomal large subunit biogenesis (GO:0042273)&			3	5	13	30	32	34	47	57	77	82\\
ribosomal small subunit biogenesis (GO:0042274)&			3	5	13	32	34	42	47	57	77	82\\
ribosomal subunit export from nucleus (GO:0000054)&			13	34	47	82\\
ribosome assembly (GO:0042255)&			34	47	82\\
RNA modification (GO:0009451)&			5	13	32	34	47	57	77	82\\
rRNA processing (GO:0006364)&			3	5	13	14	30	32	34	42	47	50	57	77	82	88\\
sporulation (GO:0043934)	&		12	17	28	46	48	72	85	90\\
transcription from RNA polymerase I promoter (GO:0006360)&			34\\
translational elongation (GO:0006414)&			31	35\\
\hline
\end{tabular}
\label{tab:GO}
\end{table}

\subsection*{Latent Variable Conditional Mean Estimation}
Because many of the latent nodes are associated with the same TFs, we examine how their activities vary across the biological
conditions to establish unique roles for latent nodes that initially appear to be redundant.
We employ Gaussian belief propagation, which is computationally efficient in the latent tree model, to infer
the conditional means for LNs under all the $498$ conditions present in the dataset (Figure~\ref{Fig:LNActivity}a).  The conditional means can be thought of as the
signed activity of a latent node in each condition.  The associations between TFs and latent nodes can be used
to transfer the latent node activities to the corresponding TFs.  For example, Fhl1 significantly overlaps with
nine latent nodes (Figure~\ref{Fig:LNActivity}b), and the activity levels of these latent nodes are shown in
Figure~\ref{Fig:LNActivity}c.  Although there is a common general trend in the activity profiles of all nine
latent nodes --- negative activity in conditions to the left and positive activity in conditions to the right ---
there are obvious differences in the conditional means of these latent nodes as well.  Figure~\ref{Fig:LNActivity}c
highlights one group of conditions where this is particularly evident, which is shown in greater detail in Figure~\ref{Fig:LNActivity}d.
In these conditions, which are primarily late time points in nitrogen depletion and stationary phase, four Fhl1-associated
latent nodes show positive activity and the other five show little or negative activity.  Although the latent nodes
appear similar because their gene neighborhoods all overlap with Fhl1-bound genes, they in fact capture different aspects
of Fhl1 regulation.

\begin{figure}[htp]
    \centering
    \includegraphics[width=\textwidth]{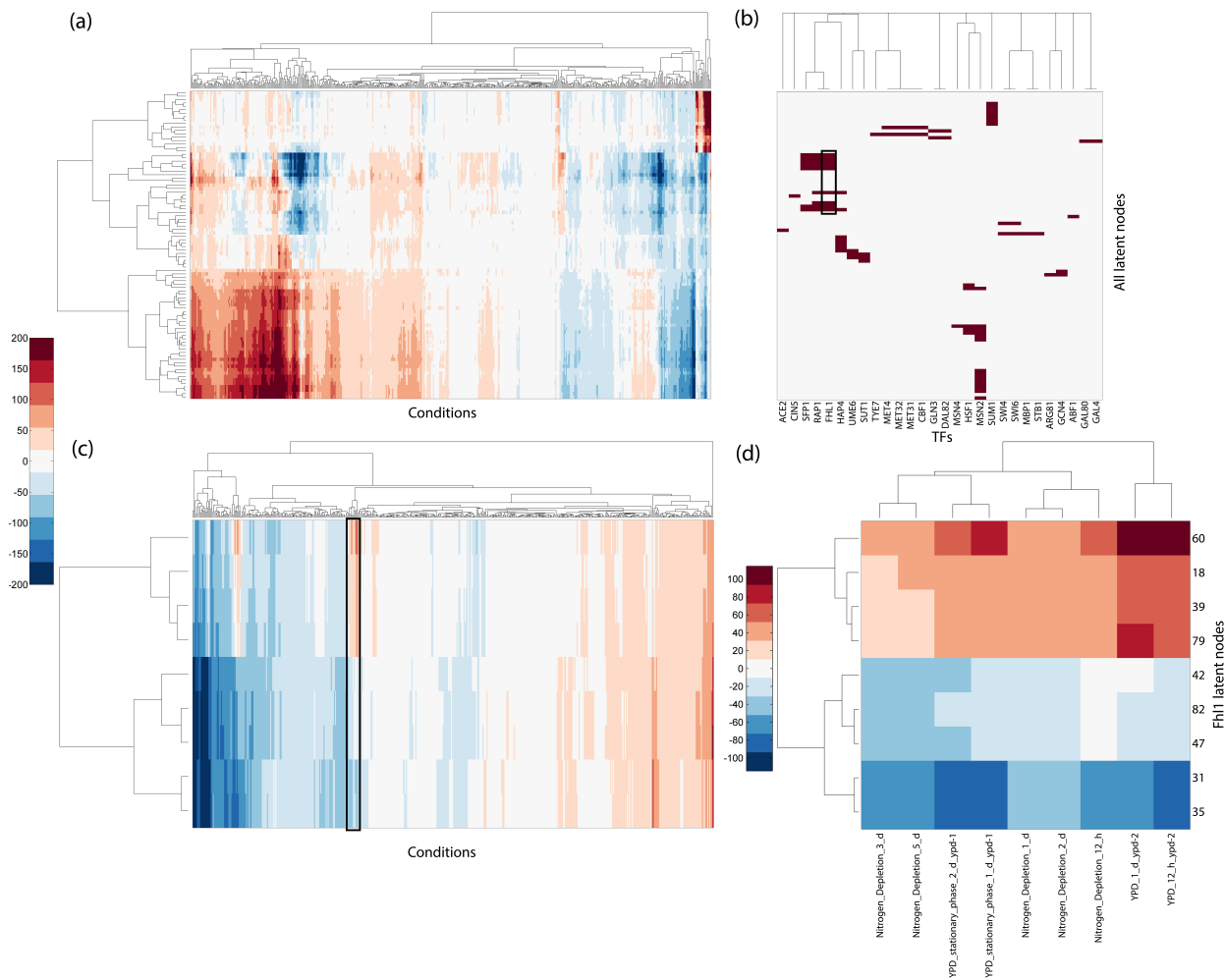}
    \caption{(a) A hierarchical clustering of the latent node activities and biological conditions.
    Red denotes positive regulatory influence and blue reflects inhibition.  (b) The mapping of TFs and latent nodes (as in Figure~\ref{Fig:BipartiteTFLV})
    reveals TF activity across the conditions.  The latent node activities can be transferred to the corresponding TFs.  The black box
    highlights latent nodes mapped to Fhl1.  (c) The activity levels of the nine latent nodes that map to Fhl1.  Fhl1 shows distinct modes of regulation.  It
    functions primarily as a repressor in conditions to the left and as an activator in conditions to the right.  Columns do not appear in the same
    order as in (a).  In some conditions, such as those outlined in the black box, the different latent nodes mapped to Fhl1 capture distinct
    activity levels.  (d)  The highlighted conditions from (c), which are mostly nitrogen depletion and stationary phase, demonstrate how latent nodes
    that correspond to the same TF can represent unique activity profiles.  Rows are in the same order as in (c).}
    \label{Fig:LNActivity}
\end{figure}

\subsection*{Combinatorial Regulation}
Transcription factors do not act in isolation, but rather participate in complex interactions with other TFs such
as cooperative regulation~\cite{ravasi2010atlas}, competitive DNA binding~\cite{wasson2009ensemble}, and functional redundancy~\cite{gitter2009backup}.
Our mapping between latent nodes and TFs can suggest such combinatorial relationships among TFs by revealing
groups of TFs that are associated with the same LN.  These groups of TFs potentially work together or
in competition to regulate similar genes in specific stress conditions and allow us to selectively focus on $30$ pairs of TFs
that are likely to function jointly from among all $4560$ TF-TF pairs.

Our filtering strategy effectively identifies pairs of TFs having the capacity to operate jointly.
We used BioGRID~\cite{chatr2013biogrid} to confirm whether there are known relationships between the TFs
we predict to function cooperatively or competitively.  Many of these TF pairs interact physically, supporting their putative combinatorial regulation (Figure~\ref{Fig:BipartiteTFLV}, black edges).  For instance, we predict a relationship between Gal4 and Gal80, which are
both associated with latent node $53$.  Gal80 has been shown to bind to Gal4's activation domain in a galactose-dependent manner,
which inhibits Gal4's transcriptional activity~\cite{traven2006}.
Other TF pairs exhibit various
types of genetic interactions (Figure~\ref{Fig:BipartiteTFLV}, red edges), which can indicate cooperative or redundant
relationships depending on the specific type of genetic interaction.  Furthermore, when we consider
indirect physical and genetic interactions in which the two TFs do not directly interact but interact
with a common third protein, we find that all of the TF-TF pairs we identified have some form of
known physical or functional relationship (Figure~\ref{Fig:BipartiteTFLV}, dashed edges).

To further probe TF-TF relationships, we
study the DNA binding motifs \cite{macisaac2006improved} of TFs that are associated with the same LN.  Some groups of TFs, such as Arg81-Gcn4
and Met4-Met31-Met32, have very similar binding motifs suggesting the presence of a multi-TF regulatory mechanism.
The exploration of the exact nature of that relationship (competitive, redundant, or cooperative)
can be guided by the latent tree, which provides the specific conditions in which these TFs are expected to be active
together.

\subsection*{Motif Enrichment of Unannotated Latent Nodes}
Our previous association of TFs and latent nodes depended on TF-gene interaction data, which are not available
for many species and cell types in higher-order organisms.  However, even without TF-gene interactions it is
still possible to annotate latent nodes.  We can perform \emph{de novo} motif discovery on the
gene neighborhood of each latent node, searching for common DNA sequences in the genes' promoter regions that
may be bound by the hidden regulator.  Given significant motifs, we can scan motif databases to identify
TFs with a known binding motif that matches the \emph{de novo} motifs.

We demonstrate this general strategy by searching for regulators whose activity could be represented by
the $39$ latent nodes that do not significantly overlap any TFs
(Supplementary Table~\ref{STab:TFSignificance}).  The WebMOTIFS~\cite{romer2007webmotifs} motif discovery tool identifies
significant motifs for $35$ latent nodes.  These latent node neighborhoods are enriched for
one to three motifs, except latent node $43$, which has ten distinct motifs.  We use STAMP~\cite{mahony2007stamp}
to align the discovered motifs to known yeast TF binding motifs in the JASPAR~\cite{portales2010jaspar} database
and find strong matches for all \emph{de novo} motifs (Supplementary Table~\ref{STab:DeNovoMotifs}).

Nine of the TFs --- Cup9, Gcr1, Ime1, Ndt80, Nhp10, Rei1, Rgm1, Stb3, and Tod9 ---
that are the best match for a \emph{de novo} motif are absent from the TF-gene interaction dataset\cite{macisaac2006improved}.  The other
$14$ are represented in the TF binding interactions but do not exhibit significant overlaps with these
latent nodes.  One explanation is that these are lower confidence TF-LN associations.  Alternatively, these associations
may have been missed previously
because the TF-gene interaction dataset represents a limited number of experimental conditions (typically rich medium).
Yeast TFs can change their binding patterns in different conditions~\cite{harbison2004}
so it is possible that the genes in a latent node neighborhood
are bound by a TF in the stress conditions in which the latent node is active but not in normal growth
conditions.  This could explain why Msn4 only annotates a single latent node when using the TF-gene binding
data despite its role as a primary general stress response TF.  Msn4 has been characterized as a `condition-altered' regulator~\cite{harbison2004}
(it binds different, but partially overlapping, groups of genes in different conditions),
and it is the best matching TF for nine latent nodes
in our motif discovery analysis.  Thus, not only can \emph{de novo} motif discovery be used to annotate latent trees
in settings where TF-gene interactions are not available, but it can also compensate for TF associations that are missed
due to the condition-specific nature of TF binding.

\subsection*{Comparison with ARACNE}
Algorithms that only model observed gene expression variables and implicitly assume that gene expression is representative of regulator activity recover a fundamentally different type of network structure than the latent tree graphical model.
To illustrate, we compare with ARACNE~\cite{margolin2006aracne,margolin2006aracneprot}, a regulatory network inference algorithm that performed comparably with the best methods on the \emph{S. cerevisiae} dataset in the DREAM5 network inference challenge~\cite{marbach2012wisdom}.  Briefly, it utilizes the pairwise mutual information between genes' expression values to learn a general network structure (in contrast with a tree) using the data processing inequality.
ARACNE does not distinguish between genes and TFs.  Each TF is represented by the expression levels of the gene that encodes it.  Hidden nodes are not incorporated in ARACNE so it cannot recover the true conditional independence structure when latent effects are present, as they typically are in biological settings.

We run ARACNE using the same yeast gene expression data that has been filtered to remove genes that do not covary with any other genes (Methods). The ARACNE network contains only nine TFs; all other TFs did not vary substantially across the biological conditions.  This reinforces the limitations of assuming that TF activity is coupled with TF expression.  Our latent tree analysis (Figure~\ref{Fig:BipartiteTFLV}) and prior research into the conditions targeted in the expression dataset provide strong evidence that far more than nine TFs are involved in regulating these processes.  Alternative strategies for filtering the expression data and other network reconstruction algorithms could recover additional TFs, but this coarse comparison demonstrates the distinct assumptions and goals of the latent tree model and traditional methods.

\section*{Discussion}
The latent tree graphical model is a powerful unsupervised technique for recovering hidden structure in
gene expression compendia and detects different types of gene expression
relationships than traditional regulatory network reconstruction
techniques.  Our study of stress response and non-stress conditions
in yeast reveals $90$ latent factors that explain gene co-expression.  Many of these latent nodes
represent specific TFs or groups of TFs, and the latent tree model
recovers the condition-specific activities of these TFs.  \emph{De novo} motif discovery
reveals additional TF-latent node associations, many involving the general stress transcription
factor Msn4, that were missed due to condition-altered TF binding.
We further demonstrate that the latent tree's predictions are not captured by an algorithm that assumes
gene expression is representative of TF activity.

Latent variable models have been successfully applied in a variety of gene expression
analyses.  Surrogate Variable Analysis~\cite{leek_capturing_2007,leek_sva_2012} accounts
for unobserved confounding factors, and CellCODE~\cite{chikina_cellcode_2015}
models variation in cell type composition during differential expression analysis.
Analogous to principal component analysis (PCA), latent variable methods can recover informative low-dimensional
representations of high-dimensional expression data~\cite{he_learning_2012,chen_consistent_2015}.
Autoencoders trained on gene expression data~\cite{tan_unsupervised_2014,chen_learning_2016} provide
an alternative compact representation, and the hidden units may be associated with transcription factors
and biological pathways or attributes such as disease status and patient survival in a clinical setting.
The Latent Differential Graphical Model~\cite{tian_identifying_2016} identifies rewiring in gene
regulatory networks from two conditions, and BicMix~\cite{gao_context_2016} can detect subsets
of genes that are co-regulated only in specific conditions.

Early methods that incorporated latent variables in regulatory network modeling were more
restricted than our latent tree graphical model.
Bayesian networks with latent variables were applied to study the yeast galactose regulatory
network but only considered a limited number of genes~\cite{hartemink2001graphical,yoo_discovery_2002}.
Another graphical model~\cite{zhang2012learning} assumed that gene expression represents TF activity and only included hidden variables to model potentially unknown TFs and confounding factors.
More recently, INSPIRE~\cite{celik_extracting_2016} developed a latent variable Gaussian graphical model
for a compelling study of gene expression in ovarian cancer.  INSIPRE is unique in its ability
to combine gene expression samples that contain different genes,
but some of the underlying motivations are similar to our latent tree graphical model.  INSPIRE represents
gene modules with latent variables, learns conditional dependencies among the module latent variables, and assigns each
gene to exactly one module.  Unlike INSPIRE, our latent tree approach allows direct gene-gene edges in the graph
but is restricted to tree topologies.  In the latent tree, genes may belong to multiple modules, which provides additional
flexibility, but the modules are not provided by the graph structure.  Rather, we must define the
extended neighborhood of influence for each latent node in a post-processing step.  Directly comparing
INSPIRE and the latent tree graphical model is an important topic for future work.

Numerous methods cluster genes based on their expression levels.
Given a cluster of genes, it is possible to find putative regulators
using the same type of TF enrichment analysis we performed to annotate the latent
nodes in our latent tree model.  However, most clustering approaches cannot
model the dependencies among genes and regulators or recover
the TF activity levels across the biological conditions, which requires
inference in the graphical model.  The inferred regulator activities can
reveal biological phenomena, such as our prediction that Fhl1-associated
latent nodes can exhibit both positive and negative activity in some
stress conditions (Figure~\ref{Fig:LNActivity}d).

Many network reconstruction algorithms integrate gene
expression and regulator binding (from DNA binding motifs, ChIP-chip, ChIP-seq,
analysis of epigenetic features, or other similar sources) and derive regulator
activities from the expression levels of the bound targets.
These methods work well when the experimental conditions (for example, cell type and environment)
of the regulator binding data match the gene expression conditions.
However, when species- and condition-specific binding data are unavailable, these methods are
either inapplicable or can miss important regulators, as we observed with the condition-altered
binding of Msn4.
When appropriate regulator binding interactions are available, they provide a natural way
to interpret some of the latent nodes in the latent tree model.  Even in these cases
the latent tree can still provide a complementary perspective about gene expression.
By first learning the latent tree structure and then annotating latent nodes in an optional
post-processing step, the latent tree model can detect other unobserved factors that induce
dependencies among co-expressed genes without being biased by regulator binding interactions.

In contrast to algorithms that assume TF
activity is accurately represented by gene expression, our latent tree algorithm infers
the activities of the hidden regulators without using
TF expression.  Previous studies have shown that mRNA expression is not always a reliable
proxy for protein abundance or activity due to post-transcriptional regulation and other
effects~\cite{nagashima2008phosphoproteome,ghazalpour2011comparative,waters2012network,
osmanbeyoglu_linking_2014,stefan_inference_2015,arrieta-ortiz_experimentally_2015,
oconnell_simultaneous_2016}.
The regulatory activity of a protein is a function of its protein abundance,
sub-cellular location, post-translational modifications, and other factors, which makes it
difficult to observe and poorly approximated by gene expression.  Consequently,
algorithms that tie regulator activity to expression can miss important TFs that
are not differentially expressed.  Indeed, the DREAM5 network inference challenge
revealed that algorithms that assume mRNA levels of TFs and their target genes are mutually
dependent can successfully reconstruct \emph{in silico} regulatory networks (where this assumption
holds) but perform poorly on expression data from a eukaryotic organism like yeast (where the assumption
fails)~\cite{marbach2012wisdom}.

Our model's predicted osmotic stress regulators
illustrate the advantages of decoupling TF expression and regulator activity.
The primary drivers of the transcriptional response to hyperosmotic stress have been well-studied
computationally~\cite{gitter2013linking,chasman_pathway_2014} and experimentally and include the TFs Hot1, Msn2, Msn4, and Sko1~\cite{capaldi2008structure}.  We identified the LNs that are most active in the osmotic stress conditions in our expression dataset and used the
TF-LN bipartite graph to recover the TFs that those LNs represent (Figure~\ref{Fig:BipartiteTFLV}).
Although Msn2 and Msn4 only display at most
$2.2$- and $1.7$-fold increases in expression, respectively, across the $14$ osmotic stress samples and decreases in expression of similar magnitudes, the latent tree correctly recovers them as osmotic stress regulators.
Hsf1 is also predicted to be an osmotic stress regulator, most likely because some of the osmotic stress experiments were performed
simultaneously with a mild heat shock~\cite{gasch2000genomic}, and Hsf1 controls transcription in heat shock response.
Hot1 may be represented by a latent node, but was not annotated because it is not present in the TF binding data\cite{macisaac2006improved}.
The Module Networks method~\cite{segal2003module}, which uses regulator expression
to construct condition-specific regulatory modules, predicts seven regulators that control its
`Energy and Osmotic stress' modules~\cite{Segal:Module:Online}.  However, Module Networks
fails to recover any of these four core osmotic stress TFs because their expression does not
correlate strongly enough with the expression profiles of the genes that respond to this
stress.

Representing the drivers of transcriptional regulation as latent variables instead of
observed variables whose activity is approximated by gene expression could also guide the
discovery of other classes of regulators.  It is well-understood that RNA levels are not controlled
by TFs alone.  MicroRNAs are recognized
as an important type of expression regulator~\cite{pasquinelli2012micrornas}
with relevance to human disease, and competitive RNA binding
can indirectly influence mRNA expression in some cases~\cite{thomson_endogenous_2016}.  Although
the latent tree model cannot directly identify these novel mechanisms, recognizing that
some latent variables do not correspond to known transcriptional regulators provides a
direction for further investigation.

Our current analysis focuses on transcriptional regulation in yeast because the vast
collection of previous studies allows us to verify many of the latent tree model's
predictions.
Having established the utility of the latent tree approach, future work could include
applying the model to provide insights into other species and human disease.
Transcriptional regulation in plants is poorly understood relative to yeast, but hundreds of
gene expression samples are publicly available for plant species --- including crops such as
barley, grape, maize, rice, soybean, tomato, and wheat --- and can be used to estimate
gene co-expression~\cite{yim2013planex}.  This allows us to study transcriptional regulation
in these crops using our latent tree approach, and plant TF databases such as PlantTFDB~\cite{zhang2011planttfdb}
could be used to annotate the latent nodes.

In humans, efforts to profile
cancer cell lines~\cite{barretina2012cancer} and primary tumors~\cite{tcga2012comprehensive}
have yielded large-scale expression datasets.  Despite the lack of a comprehensive map of
TF binding in specific types of cancer cells, latent nodes could be annotated using the same
motif discovery approach we described for yeast or by integrating predicted TF binding
interactions derived from epigenetic features~\cite{neph2012expansive} and
microRNA binding~\cite{friedman2009most}.  The latent variables could also generate insights
into cancer phenotypes.  INSPIRE demonstrated that latent variables
learned with an unsupervised algorithm can be more predictive of histological and clinical phenotypes
such as stroma type, patient survival, and tumor resectability than the expression levels of
all genes~\cite{celik_extracting_2016}.  The latent tree algorithm is not limited
to gene expression data and could be used to probe the relationships between mutations,
copy number variation, protein abundance, and phenotypic data available in the cancer
collections as well.

\section*{Methods}
\subsection*{Unsupervised Learning of Latent Tree Models}
\label{sec:structure_parameter}
Learning latent tree  models involves discovering structural relationships (structure learning) and estimating the strength of such relationships (parameter estimation).  Many methods have been developed previously for learning the structure of latent trees~\cite{erdos99,mossel2007distorted,Choi&etal:10JMLR}, and we employ the method of Choi et al.~\cite{Choi&etal:10JMLR}. Parameter estimation is carried out through the standard expectation maximization (EM)~\cite{dempster1977maximum} procedure. Below, we introduce graphical models for Gaussian distributions and then describe the main steps of the latent tree method of Choi et al.~\cite{Choi&etal:10JMLR}.

\subsubsection*{Gaussian Graphical Models}
A Gaussian graphical model is a family of jointly Gaussian distributions that factor in accordance
with a given graph.  Let $V$ represent the node set and $E$ the edge set. Given a graph $\mathcal{G} = (V,E)$,
we consider a vector of Gaussian
random variables $X = [X_1,X_2,...,X_p]^T$ where each node $i \in V$ is associated with a scalar Gaussian
random variable $X_i$ and $p = \left|V\right|$.
A Gaussian graphical model Markov on $\mathcal{G}$ has a probability density function
(pdf) that is parameterized as

\begin{equation}
f_X(x)\propto exp\left[\frac{1}{2} x^T J_\mathcal{G} x + h^Tx\right],
\end{equation}
where $J_\mathcal{G}$ is a positive-definite symmetric matrix whose sparsity pattern corresponds to that of the graph $\mathcal{G}$. More precisely, $J_\mathcal{G}(i,j)=0 \Longleftrightarrow (i,j)\notin E$. The matrix $J_\mathcal{G}$ is known as the potential or information matrix with the non-zero entries $J(i,j)$ as the edge potentials, and the vector $h$ is the potential vector.
This form of parameterization is known as the information form and is related to the standard mean-covariance parameterization of the Gaussian distribution as
$
\mu=J^{-1}h,\  \Sigma=J^{-1},
$
where $\mu:=\mathbb{E}[X]$ is the mean vector and $\Sigma:=\mathbb{E}\left[(X-\mu)(X-\mu)^{T}\right]$ is the covariance matrix.

\subsubsection*{Additive Metric in Tree Graphical Models}
For Gaussian models, the information distance between any two nodes $i$ and $j$ in a tree $T$ is defined as \beq d_{ij}:=-\log |\rho_{i,j}|, \eeq where $\rho_{i,j}:= \frac{Cov(X_i,X_j)}{\sigma_{X_i} \sigma_{X_j}}$ denotes the correlation coefficient between nodes $i$ and $j$. Note that $Cov(\cdot)$ is the covariance and $\sigma$ is the standard deviation. These distances $\{d_{ij}\}$ can thus be estimated using the samples corresponding to the observed nodes.  If the joint probability distribution is a tree-structured graphical model, then the information distances are additive along paths in the tree as
 \begin{equation}
d_{kl}=  \sum\limits_{(i,j)\in \text{Path}(k,l)}d_{ij}
 \end{equation}
Learning a latent tree  can thus be reformulated as learning a tree structure $T$ given pairwise (estimated) distances $\bfd:=\{\hd_{ij}:i,j\in V\}$ between the observed nodes $i$ and $j$, $\forall i,j \in \mathcal{G}$.

\subsubsection*{Sibling Grouping and Recursive Grouping}
The latent tree algorithm constructs a tree in a bottom-up manner. Every node (except the root) has exactly one parent in the tree. We call a group of nodes siblings if they share the same parent. The algorithm first classifies nodes that are under consideration at the current iteration into sibling or parent-child groups~\cite{Choi&etal:10JMLR}.
A sibling test is conducted to ascertain which nodes under consideration are siblings. A family is a group of nodes that are either leaf-siblings or a group of leaf-siblings with their common parent.  We then eliminate the nodes that are in families of more than one member. Let $\Delta_{ijk} := d_{ik}-d_{jk}, \ \forall k \in V\backslash \{i,j\}$ denote the difference between two information distances $d_{ik}$ and $d_{jk}$. We see that if $\Delta_{ijk} \equiv C , \ \forall k \in V\backslash \{i,j\}$, where $C$ is some constant, then $i$ and $j$ belong to the same family. Furthermore, if $C=d_{ij}$ , we know that $i$ is a leaf and $j$ is its parent.

After the sibling test, we identify nodes that belong to the same family.
Latent nodes are introduced when a family of nodes does not contain an observed parent.  First, we group all observed nodes into families. Four scenarios may happen: (1) a single node family, (2) a leaf node with a parent, (3) siblings with a parent, and (4) siblings with no parent. The algorithm introduces a new hidden parent in the fourth scenario and removes the current lowest layer nodes from further consideration, fixing this portion of the tree structure. This procedure is repeated  recursively and is termed the recursive grouping procedure (Figure~\ref{Fig:LatentTreeLearning}).

\begin{figure}[htp]
  \centering
  \def\svgwidth{6.9in}
	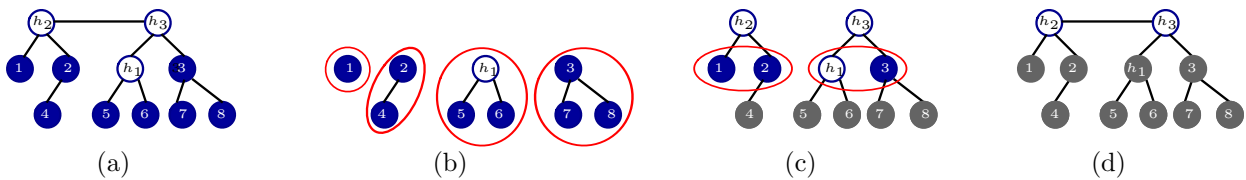
  \caption{An example of the recursive grouping algorithm to learn the latent tree structure. (a) The ground truth latent tree that we would like to recover. We only observe distances between observed nodes (blue). (b) After grouping the observed variables into families (red curves), the first latent variable is introduced. (c) A portion of the latent tree structure is fixed (grey nodes) and the family identification step is repeated. (d) The latent tree learned by the recursive grouping algorithm.}
	\label{Fig:LatentTreeLearning}
\end{figure}

The latent tree structure learning algorithm recovers accurate models even when applied to noisy biological data.  It can be verified from the Central Limit Theorem and continuity arguments~\cite{Choi&etal:10JMLR} that the gap between the true ($d_{ij}$) and estimated ($\widehat{d}_{ij}$) distances between observable nodes $i$ and $j$ is bounded as
\begin{equation}
\widehat{d}_{ij}-d_{ij}=O(n^{-1/2}),
\end{equation}
where $n$ is the number of samples.

\subsubsection*{CL-grouping Procedure}
The recursive grouping algorithm requires a large number of sibling tests and thus is not computationally efficient. Moreover, merging the results of all these tests can lead to error accumulation.  A more efficient alternative is the $\clgroup$ method (Chow-Liu Grouping)~\cite{Choi&etal:10JMLR}. This method recursively modifies the estimate of the latent nodes by operating on local neighborhoods and adding new latent nodes. In the initial step, the method constructs a minimum spanning tree $\mst(V;\bfd)$ (also known as the Chow-Liu Tree) over the observed nodes (gene nodes in this context) using distances $\bfd$. This global step groups the observed nodes that are likely to be close
to each other in the true underlying latent tree, thereby guiding subsequent applications of the recursive grouping algorithm. Specifically, the recursive grouping procedure recursively conducts sibling tests in local neighborhoods of the $\mst(V;\bfd)$ in each iteration.  Furthermore, if adding latent variables in a local neighborhood does not improve the Bayesian Information Criterion, which is a penalized likelihood score, then those latent variables are rejected. Because the sibling tests are limited to local neighborhoods, the learning procedure is efficient.

\subsubsection*{Parameter Tuning and Contraction of Latent Nodes}
In practice, biological data are noisy and the criteria for the sibling tests need to be relaxed because the estimated distances are imperfect.  As discussed above, the family identification test between $i$ and $j$ requires examining estimated $\widehat{\Delta}_{ijk}$ for all $k\in V\backslash \{i,j\}$.  If we allow for relaxation, then
\begin{equation}
\widehat{\Delta}_{ijk}  \equiv C + \epsilon_1
\end{equation}
is the relaxed condition for the family test,
and
\begin{equation}
C = \widehat{d}_{ij} + \epsilon_2
\end{equation}
is the relaxed condition for the parent-child test.  $\epsilon_1$ are $\epsilon_2$ are relaxation parameters.
$k$-means clustering using the silhouette method is employed to select the relaxation parameters for the sibling tests.

In addition, a contraction step removes latent nodes that have information distance smaller than some predefined threshold to some observed node(s).  In our experiments, latent nodes are contracted if the information distance is less than $0.9365$.  We select this threshold because it produces a latent tree with $90$ latent nodes, which is similar to the number of yeast TFs in the TF-gene binding dataset we use to annotate the latent tree\cite{macisaac2006improved}.  General strategies for controlling model complexity in graphical models are beyond the scope of this work.  However, when there is no prior expectation for the number of latent nodes, the latent node activity signatures (as shown in Figure~\ref{Fig:LNActivity}a) can guide the choice of the contraction parameter.  If there are many latent nodes with similar activity signatures, then the contraction step may be beneficial for controlling redundancy in addition to the Bayesian Information Criterion regularization that is already part of the structure learning.

After learning the structure and parameters of the latent tree model, we perform Gaussian belief propagation~\cite{Yedidia2003jan1} on the tree to obtain the conditional means for LNs conditioned on the samples of the observed gene nodes.   In order to estimate these conditional means, we use the signed correlation coefficients, whereas the information distances calculated for structure learning use absolute values of the correlation coefficients.

\subsection*{Data}
We downloaded microarray expression data from~\cite{GaschData}, which consists of $498$ samples including many
stress response conditions as well as other
experiments such as cell cycle (data compiled from~\cite{gasch2000genomic,gasch2001genomic,lyons2000genome}
and other studies)\footnote{The dataset is not longer available from the original website but can be accessed from the Internet Archive at\\
\url{https://web.archive.org/web/20130629025218/http://gasch.genetics.wisc.edu/datasets.html}}.  We imputed the expression data to account for missing data (Supplementary Methods).
For the TF-gene binding data we used only interactions
with $p$-value less than $0.001$ and binding motifs conserved in at least two other yeast
species\cite{macisaac2006improved}.
We filtered these TF-gene interactions to remove gene targets that are not expressed and retain $96$ TFs that bind at least one expressed gene.
We obtained physical protein-protein and genetic interactions from BioGRID version $3.2.96$\cite{chatr2013biogrid} and
removed all self-interactions and interactions involving non-yeast proteins.  To identify
indirect TF-TF interactions we searched for pairs of TFs that do not interact directly but have a
common neighbor in the interaction network.  We excluded physical-genetic interactions where the
neighbor has a protein-protein interaction with one TF and a genetic interaction with the other.

\subsection*{Gene Selection}
In large-scale genome-wide expression data, the number of genes is usually much larger than the number of samples. Dimensionality reduction or feature selection is beneficial when analyzing such datasets. Traditional methods such as PCA select principle components for dimensionality reduction, but this  leads to a loss of information from minor components. In our work, PCA is not used to select features because there is no biological reason to believe the minor components are unimportant. Instead, we focus on genes that exhibit condition-specific expression changes because these genes enable us to recover condition-specific regulatory modules.
Specifically, we compute the covariance for all gene-gene pairs and remove a gene if its maximum covariance with all other genes is less than $0.8683$. We choose this threshold because it selects approximately $1000$ covarying genes. After filtering, $1035$ genes are retained out of all $5998$ genes.

\subsection*{Annotating LNs with TFs and GO Terms}
The TF binding data are used to annotate the LNs of the latent tree.
We search for a TF that binds many of the gene nodes that are in the neighborhood of a latent node, which suggests that the TF may cause the gene nodes' expression levels to be correlated across the samples.  We define an extended neighborhood of influence of a LN and compare it to the set of genes bound by a TF to see if there is a statistically significant overlap between these two groups of genes.

The extended neighborhood of influence for each LN consists of the set of genes that are highly correlated with that latent node. More precisely, a distance matrix $Dist \in \mathds{R}^{k\times p}$ is calculated with each entry representing
\begin{equation}
Dist_{i,j}:=-\log(\left|\rho (y_i, g_j) \right|),
\end{equation}
where $k$ is the number of latent variables, $p$ is the number of observed gene variables, and $\rho(\cdot)$ is the correlation coefficient between latent node $y_i$ and gene $g_j$. Thus a smaller information distance implies a higher correlation.

The threshold to select the extended neighborhoods of influence for latent nodes is dynamically tuned. We define $d_{\text{min}}:=\min\limits_{i,j} {Dist}_{i,j}$ and $d_{\text{max}}:=\max\limits_{i,j} {Dist}_{i,j}$ and select the neighbors of latent variable $y_i$ \begin{equation}
\mathcal{N}(y_i):=\left\{ { g_j: Dist_{i,j}\le d_{\text{min}}+\lambda\left( d_{\text{max}}-d_{\text{min}}\right)  } \right\},
\end{equation}
where $\lambda$ is a tunable parameter. Smaller $\lambda$ leads to more stringent neighborhood selection. In our experiments, we set $\lambda = 0.15$.

We compare the extended neighborhood of influence of a LN with the set of genes bound by a TF to see if the overlap is statistically significant. We use Fisher's exact test~\cite{fisher1922interpretation} to test all TF-LN pairs using the $1035$ genes and control the FDR under $0.05$. Benjamini–Yekutieli's\cite{benjamini2001control} method can be used to adjust raw $p$-values and obtain the decision rule based on the adjusted $p$-values, thus controlling the FDR. However, Benjamini–Yekutieli's method assumes positive regressive correlations, which need not hold in our setting. Instead, we employ a Bayesian approach to estimate the FDR under arbitrary correlations using the fdrtool R package\cite{strimmer2008fdrtool}.  This approach can directly calculate FDR instead of thresholding on the adjusted $p$-values (Supplementary Methods).

We annotate latent nodes with GO biological process terms using the same statistical test described above for associating TFs and latent nodes and the same
FDR threshold of $0.05$.  We downloaded yeast GO Slim mappings from the \emph{Saccharomyces} Genome Database~\cite{GOSlim,cherry_saccharomyces_2012} on May $9$, $2013$
and retained only the biological process terms.

\subsection*{Motif Enrichment}
We performed \emph{de novo} motif discovery with WebMOTIFS~\cite{romer2007webmotifs}.  For each
latent node that did not overlap with any TFs, we searched for significant motifs in
its extended neighborhood using the default WebMOTIFS settings (search sequences from -$500$ to +$200$ of the transcription
start site, expected motif length less than $12$ nucleotides, and strict significance testing).  WebMOTIFS
ran four motif discovery algorithms --- AlignACE~\cite{hughes2000computational}, MDscan~\cite{liu2002algorithm},
MEME~\cite{bailey1994fitting}, and Weeder~\cite{pavesi2004weeder} --- on each set of genes and
clustered the significant motifs.  We aligned the significant motifs to yeast motifs in JASPAR~\cite{portales2010jaspar}
using STAMP~\cite{mahony2007stamp} with the default parameters (Pearson correlation coefficient comparison
metric and ungapped Smith-Waterman alignment) but did not trim motif edges.  Supplementary Table~\ref{STab:DeNovoMotifs}
reports the discovered motifs and the top matching known TF binding motifs.  We filtered the matching motifs to
exclude non-yeast TFs.

\subsection*{Osmotic Stress Regulators}
To predict TFs that are active in the osmotic stress response, we identified $14$ samples in the expression
dataset that contain the word `sorbitol' in their name.  Sorbitol induces hyperosmotic
stress response~\cite{gasch2000genomic}.  We ranked the latent nodes by their conditional means in these
osmotic stress conditions and selected the top $50\%$ as
the active hidden regulators.  We defined the osmotic stress TFs as
all TFs that are significantly associated with these latent nodes (Figure~\ref{Fig:BipartiteTFLV}).
The osmotic stress TFs did not change when we selected only the top $40\%$ or $30\%$ of latent nodes
or restricted the analysis to the seven sorbitol samples that did not involve a
simultaneous temperature change.

\subsection*{ARACNE Comparison}
We ran ARACNE on the filtered expression dataset of $1035$ genes with geWorkbench~\cite{geworkbench},
which took approximately 8 hours on machine with a 2.9 GHz Intel Core i7 and 8 GB 1600 MHz DDR3 memory.  We
were unable to run ARACNE on the full unfiltered expression matrix in a reasonable amount of time.
We set the `Kernel Width' to be `Inferred', `p-value' to be `1.0E-7', and `DPI Tolerance' to be `0.1'.

\subsection*{Software}
The latent tree method from~\cite{Choi&etal:10JMLR} is implemented in MATLAB and is available at~\cite{latentTreeCode}.

\section*{Authors' Contributions}
AG and FH collected the data and performed the computational analysis.
FH and RV implemented the latent tree algorithm.
AG, FH, EF, and AA designed the study, analyzed the data, and wrote the manuscript.
EF and AA supervised the study.
All authors read and approved the final manuscript.

\section*{Acknowledgements}
  \ifthenelse{\boolean{publ}}{\small}{}
  FH is supported by NSF BIGDATA IIS-1251267. AA is supported in part by a Microsoft Faculty Fellowship, NSF Career Award CCF-1254106, NSF Award CCF-1219234, and ARO YIP Award W911NF-13-1-0084.  EF is supported in part by the Institute for Collaborative Biotechnologies through grant W911NF-09-0001 from the US Army Research Office (the content of the information does not necessarily reflect the position or the policy of the Government, and no official endorsement should be inferred) and by NIH grant R01-GM089903.


{\ifthenelse{\boolean{publ}}{\footnotesize}{\small}
 \bibliographystyle{bmc_article}  
  \bibliography{gitter_latent_tree} }


\ifthenelse{\boolean{publ}}{\end{multicols}}{}


\section*{Ancillary Files}
\subsection*{SupplementaryTables.xlsx}
The following tables are contained in the spreadsheet:
\begin{itemize}
    \item \suptable{STab:LatentTreeEdges} Latent tree network adjacency list.
    \item \suptable{STab:ExtNeighborhood} Latent node extended neighborhoods.
    \item \suptable{STab:TFSignificance} Significance of the TF annotations of the latent nodes.
    \item \suptable{STab:GOSignificance} Significance of the GO annotations of the latent nodes.
    \item \suptable{STab:DeNovoMotifs} \emph{De novo} motifs discovered for the unannotated latent nodes.
\end{itemize}


\section*{Supplementary Methods and Figures}
\subsection*{Missing Data}\label{Sec:Missingdata}
We use a multiple imputation method\cite{wayman2003multiple} to estimate the missing values in the yeast gene expression dataset. This strategy has been shown to reduce bias and increase efficiency compared to other ad-hoc methods such as listwise deletion and mean imputation.
Multiple imputation is a procedure for imputing multiple values for every missing data point, which results in multiple complete data sets. Those imputed values are sampled from a distribution inferred from the observed values.  After imputation with expectation maximization\cite{dempster1977maximum}, we combine the multiple results.

We assume the gene expression levels are multivariate normal $D\sim N(\mu, \Sigma)$ and that data are missing at random. Let $D_o$ denote observed data and $D_m$ the missing data. The posterior $p(\mu,\Sigma |D_o) \propto p(D_o|\mu,\Sigma)$ are the main imputation parameters we want to estimate because imputations can be made by sampling missing values from $p(D_m|D_o,\mu,\Sigma)$. We use EM to find the mode of the posterior. The R package Amelia II, a general-purpose tool for data with missing values, is used to perform multiple imputation\cite{honaker2012amelia}.  We perform imputation on the log scale gene expression data.

We checked the accuracy of the imputation by $10$-fold cross validation. The average error is $0.0015$, and the errors for all genes are depicted in Supplementary Figure~\ref{SFig:ErrorImputations}.

\subsection*{False Discovery Rate Control for Multiple Testing}
Here we describe the details of controlling for multiple hypothesis testing when associating TFs or GO terms with latent nodes.
We introduce the following definitions:
 \begin{enumerate}
 \item Multiple testing: $m$ null hypotheses $\mathcal{H}_{0i},\forall i \in [m]$ are tested simultaneously. Each $p$-value associated with $\mathcal{H}_{0i}$ is denoted as $p_i$, and $\hat{P}^m:=\{p_1,p_2,\ldots,p_m\}$ is the set of all observed $p$-values. $m_0$ of the $m$ hypotheses are the true null hypotheses and thus the proportion of true null hypotheses is $\eta_0=\frac{m_0}{m}$.
 \item Decision rule: Decision rule $\theta$ is a mapping from observations of multiple test $p$-values $\hat{P}^m$ to an action $\{ A_{\text{rej}}, A_{\overline{\text{rej}}}\}$, where $A_{\text{rej}}$ denotes the `reject' action and $A_{\overline{\text{rej}}}$ denotes the `non-reject' action.
     A decision rule with cutoff $p$-value $y_c$ is defined as
     \begin{equation}
     \theta(\hat{P}^m,y_c)=
     \left\{
     \begin{array}{ll}
     A_{\text{ rej }\mathcal{H}_{0i}} &, \forall i: p_i \le y_c\\
     A_{\overline{\text{rej}}\ \mathcal{H}_{0j}} &, \forall j: p_j >y_c
     \end{array}
     \right.
     \end{equation}
     where $A_{\text{ rej }\mathcal{H}_{0i}}$ is the action to reject the $i$-th null hypothesis, and $A_{\overline{\text{rej}}\ \mathcal{H}_{0j}}$ means do not reject the $j$-th null hypothesis.

\item False discovery rate: defined as
\begin{equation}
Fdr\left(\theta(\hat{P}^m,y_c)\right):=\Pr \left\{ A_{\text{ rej }\mathcal{H}_{0i}} | p_i\le y_c  \right\} = \frac{FP}{TP+FP}
\end{equation}
where $TP$ are the true positives and $FP$ are the false positives.
\end{enumerate}

The classic tail area-based false discovery rate (FDR) is the type of false discovery rate we focus on in this work.
The two component mixture of the observed $p$-values is
 \begin{equation}
 f(p)=
 \left\{
 \begin{array}{ll}
 \eta_0 f_0(p;\phi)&, 0\le p\le y_c \\
 (1-\eta_0)f_A(p)&, y_c\le p \le 1
 \end{array}
 \right.
 = \eta_0 f_0(p;\phi) \mathds{I}\{0\le p\le y_c\}+(1-\eta_0)f_A(p) \mathds{I}\{y_c\le p \le 1\},
 \end{equation}
where $f(p),p\in[0,1]$ is the overall $p$-value density function. Let $y_c$ denote the cutoff point under which null hypotheses with smaller $p$-values ($p(\mathcal{H}_{0m})\le y_c$) are rejected. Let $f_A(p)$ denote the alternative density with support $y_c\le p \le 1$. We assume the null density $f_0(p;\phi),0\le p \le y_c$
to be a null  model with parameter $\phi$ whose support is the `uninteresting' $p$-values (null hypotheses are rejected when corresponding $p$-values are `interesting', i.e. $p \le y_c$). Thus we have
\begin{equation}\label{Eqn:twoCompoMix}
F(p)=\eta_0F_0(p;\phi)\mathds{I}\{0\le p\le y_c\}+(1-\eta_0)F_A(p)\mathds{I}\{y_c\le p \le 1\},
\end{equation}
where $F(p)$ is the cumulative distribution function associated with the marginal density $f(p)$.
Note that $\eta_0$ is the null proportion that we are to estimate.

$Fdr$ is estimated by fitting the two component FDR mixture model in Equation~\eqref{Eqn:twoCompoMix}
and estimating the cumulative distribution function $F(p)$, the proportion of true null hypotheses $\eta_0$, and $F_0(p;\phi)$ because
\begin{equation}
Fdr\left(\theta(\hat{P}^m,y_c)\right)=\eta_0\frac{1-F_0(y_c)}{1-F(y_c)}.
\end{equation}

The Grenander density estimator\cite{grenander1956theory} is used to estimate the cumulative distribution function $F(p)$, which is the decreasing piecewise-constant function equal to the slopes of the least concave majorant of the empirical cumulative distribution function\cite{strimmer2008unified}.

To estimate $\eta_0$ and $F_0(p;\phi)$, we apply a truncated maximum likelihood approach~\cite{strimmer2008fdrtool}.  Once we obtain estimates of $\eta_0$ and $F_0(p;\phi)$ we have false discovery rate control for the latent node association testing.


\begin{supfigure}[ht]
  \centering
\includegraphics[width=6in]{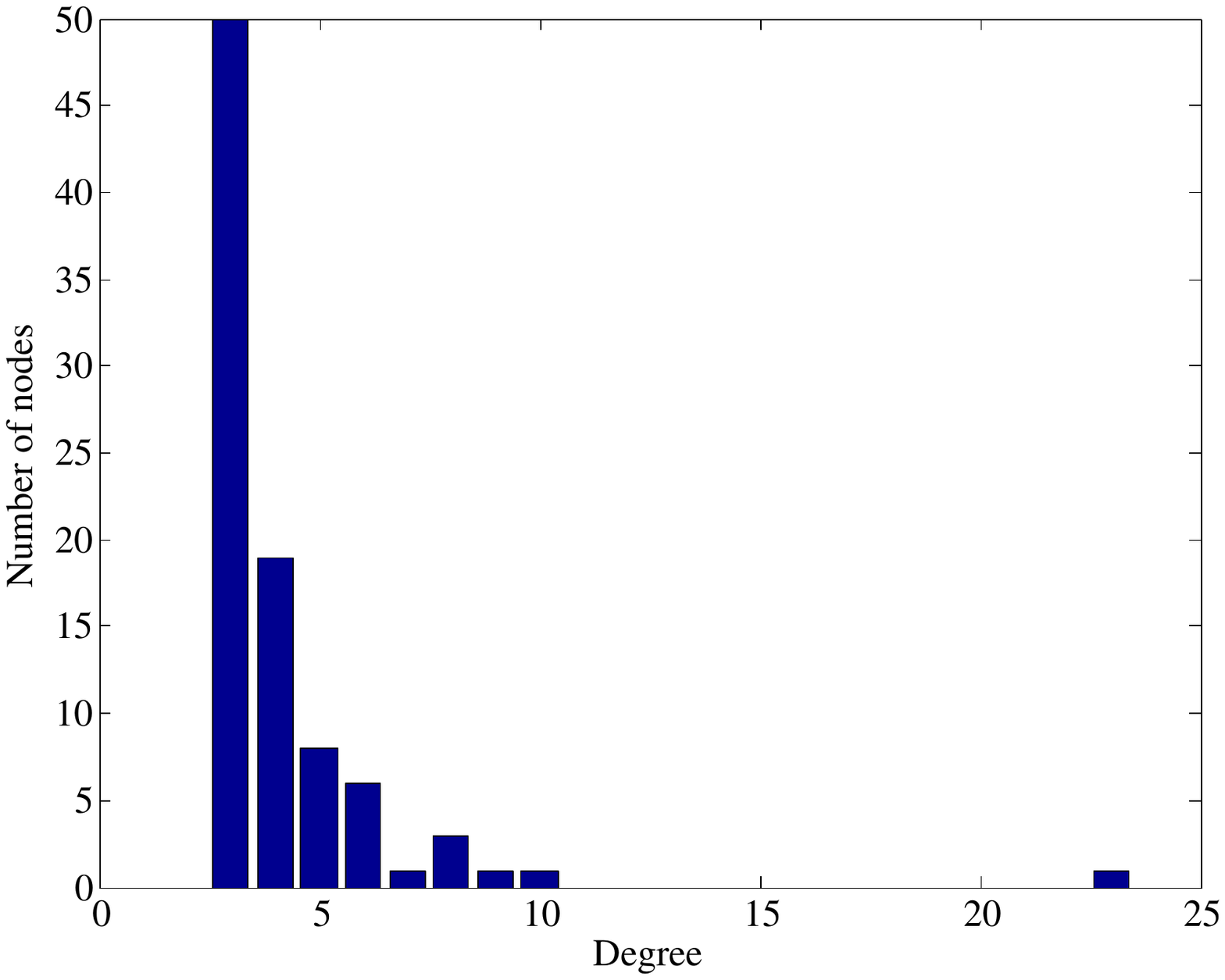}
  \caption{Degree distribution of the 90 latent nodes.
  }\label{SFig:LatentTreeDegree}
\end{supfigure}

\begin{supfigure}[ht]
  \centering
\includegraphics[width=6in]{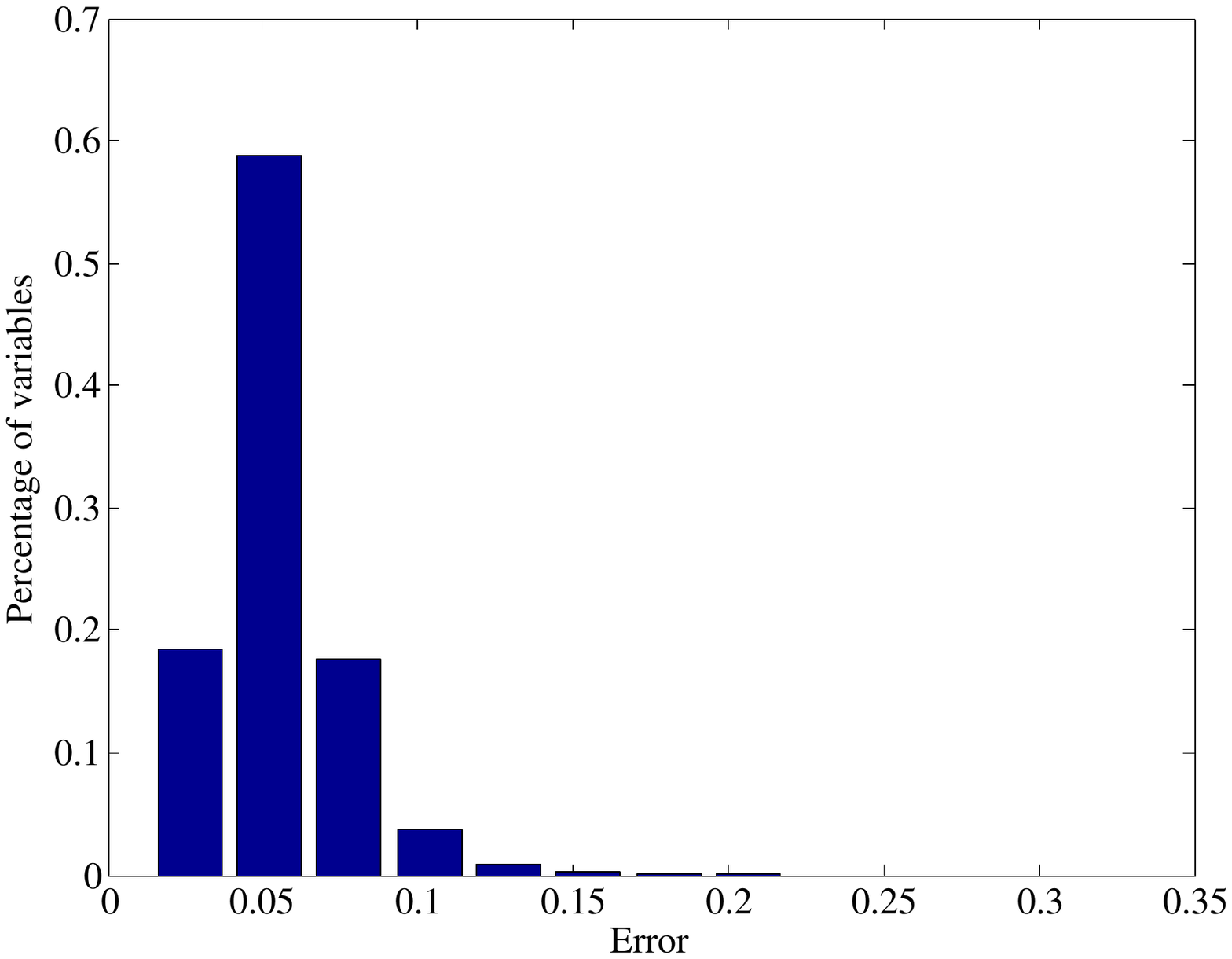}
  \caption{Gene expression imputation error via 10-fold cross validation.
  }\label{SFig:ErrorImputations}
\end{supfigure}

\end{bmcformat}
\end{document}